\documentclass[doublecol]{epl2} 
\usepackage{color}
\usepackage{ulem}

\usepackage{amssymb}

\title{ 
Dynamics of Fluid Vesicles in Flow through Structured Microchannels}

\author{Hiroshi Noguchi\inst{1,2}\thanks{E-mail:\email{noguchi@issp.u-tokyo.ac.jp}} 
\and
Gerhard Gompper\inst{1}\thanks{E-mail:\email{g.gompper@fz-juelich.de}} 
\and 
Lothar Schmid\inst{3}\thanks{E-mail:\email{lothar.schmid@physik.uni-augsburg.de}} 
\and 
Achim Wixforth\inst{3},  
\and 
Thomas Franke\inst{3}\thanks{E-mail:\email{thomas.franke@physik.uni-augsburg.de}}}
\shortauthor{H. Noguchi et al.}

\institute{
\inst{1} Institut f\"ur Festk\"orperforschung, Forschungszentrum J\"ulich, 
52425 J\"ulich, Germany, \\
\inst{2} Institute for Solid State Physics, University of Tokyo,
 Kashiwa, Chiba 277-8581, Japan, \\
\inst{3} Experimentalphysik I, Microfluidics Group, Universit\"at Augsburg, 86159 Augsburg, Germany}


\abstract{
The dynamics of fluid vesicles is studied under flow in microchannels, 
in which the width varies periodically along the channel.
Three types of flow instabilities of prolate vesicles are
found.  For small quasi-spherical vesicles -- compared to the average 
channel width -- perturbation theory predicts a transition from a 
state with orientational oscillations of a fixed prolate shape to 
a state with shape oscillations of symmetrical ellipsoidal  
or bullet-like shapes with increasing flow velocity. 
Experimentally, such orientational oscillations are observed during 
the slow migration of a vesicle towards the centerline of the channel.
For larger vesicles, mesoscale hydrodynamics simulations and experiments 
show similar symmetric shape oscillation at reduced volumes $V^* \gtrsim 0.9$.
However, for non-spherical vesicles with $V^* \lesssim 0.9$, shapes
are found with two symmetric or a single asymmetric tail.
}

\begin{document}
\maketitle

\section{Introduction}

Soft deformable objects such as liquid droplets, vesicles, and cells
show a complex behavior under flow. For example,
in simple shear flow, fluid vesicles exhibit tank-treading, tumbling, and 
swinging (also called vacillating-breathing, or trembling) motions,
depending on parameters such as shear rate, viscosity contrast, and
internal volume 
\cite{kant06,seif99,nogu04,nogu05,misb06,nogu07b,lebe08,fink08,nogu09}.
Understanding the flow behavior of lipid vesicles and red blood cells
(RBCs) 
 is not only an interesting problem of the hydrodynamics of 
deformable, thermally fluctuating membranes, but is also important 
for medical applications.  In microcirculation,
the deformation of RBCs reduces the flow resistance of microvessels.
In diseases such as sickle cell anemia, RBCs have
reduced deformability and often block microvascular flow~\cite{higg07}.  
Lipid vesicles are considered as a simple model of RBCs
and also have applications as drug-delivery systems.

The recent development of microfluidic techniques~\cite{whit06} allows
the investigation and manipulation of individual cells and vesicles, {\it e.g.}
the measurement of the dynamic pressure drop for single-cell 
deformation~\cite{abka06}, the separation of RBCs from the 
suspending plasma~\cite{yang06,abka08}, and the control of oxygen
concentration to investigate sickle cells~\cite{higg07}.
There are many potential applications such as blood diagnosis on chip.
Under steady flows in homogeneous glass capillaries and rectangular 
microchannels, vesicles~\cite{vitk04} 
and RBCs~\cite{abka08,tsuk01,pozr05,nogu05b} deform into a bullet 
(with a flattened rear end) or parachute 
(with an inside bulge at the rear end) shape.  Compared 
to these steady-flow conditions,
the vesicle dynamics in time-dependent flow is much less explored.
Only very recently, phenomena like the wrinkling of vesicles after inversion 
of an elongational flow~\cite{kant07,turi08} or shape oscillation 
of RBCs~\cite{wata06,nogu09a} and fluid vesicles~\cite{nogu09b} under 
oscillatory shear flow have been discovered.

In this paper, we propose a structured microchannel system to study
vesicle dynamics. The width of a microchannel is spatially modulated along 
the channel, so that a flowing vesicle is exposed to an oscillatory 
elongational flow. 
We have fabricated such microchannels and observe the time-dependent
vesicle deformation via optical microscopy. In parallel, we use perturbation
theory for quasi-spherical vesicles and mesoscale-hydrodynamics simulations 
for non-spherical vesicles to predict several flow instabilities. 
In particular, we predict for wider channels (compared to the vesicle size) 
a transition from shape oscillations of symmetrical bullet-like shapes to 
orientational oscillations with 
decreasing flow velocity; for narrower channels, we observe stable shapes 
with two symmetric or with a single asymmetric tail.
Our results demonstrate that structured channels are well
suited to investigate the dynamical behavior of vesicles. 

\section{Experimental Systems and Methods}

Lipid vesicles, consisting of 1,2-Dioleoyl-sn-Glycero-3-Phosphocholine
(DOPC, Avanti Polar Lipids) with $0.1$mol\% fluorescently labeled T-Red DHPE 
(Texas Red 1,2-dihexadecanoyl-sn-glycero-3-phosphoethanolamine, 
triethylammonium salt, Invitrogen, USA), were prepared 
 in aqueous sucrose solution ($200$mM) by electroformation, 
as described elsewhere~\cite{angelova2000_electroformation}.
Using this method, many giant unilamellar vesicles with diameter larger than 
$10\mu$m were produced.  Prior to the experiments, the vesicles were introduced 
into an slightly hypertonic aqueous glucose or sucrose solution. 
This procedure enables control 
of the volume-to-surface ratio by inflation and deflation of the vesicles.
The vesicles in sucrose solution are perfectly density matched. 
The density difference between sucrose and glucose solutions at $200$mM is 
$\Delta \rho/\rho = 0.012$, which implies a small buoyancy force.  

\begin{figure}
  \includegraphics[width=0.95\columnwidth]{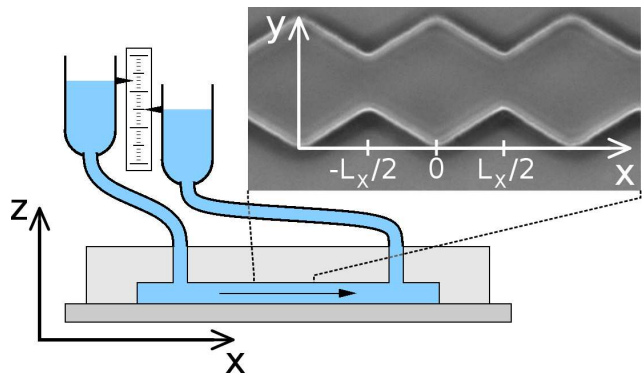}
\caption{  \label{fig:setup}
(Color online)
Side view of the channel setup with the reservoirs. The inset shows a top 
view of the modulated channel. The length of the modulation is $L_x=100 \mu$m. 
}
\end{figure}

Microchannels were fabricated by soft lithography~\cite{Whitesides1998}. 
The structure of an AutoCAD designed mask was transferred by
illumination to a photoresist (SU-8, microresist, Berlin), 
which serves as a master to cast poly(dimethylsiloxane) (PDMS) replicas. 
After treatment with an $O_{2}$-plasma, the PDMS mold was bonded onto a thin 
coverslip.  The assembly was connected to a reservoir containing the vesicle 
solution (inlet) and a water reservoir (outlet) to adjust the hydrostatic 
pressure difference, which drives the fluid through the structured 
microchannel (see Fig.~\ref{fig:setup}). All channels have a height 
$L_z=50\mu$m and a periodic length $L_x=100\mu$m. 

Due to the lift force from a wall 
\cite{olla97,abkarian2002,call08} and the migration in a parabolic 
Poiseuille 
flow~\cite{coup08,misbah2009}, vesicles are aligned close to the center of 
the channel.  In the glucose solutions, the buoyancy force leads to a 
displacement of the vesicles from the center of the channel in the 
$z$-direction (see Fig.~\ref{fig:setup}).
The balance of the buoyancy force
and the lift force \cite{olla97,call08} implies a distance
$\ell \simeq (\eta\dot{\gamma} R_0 / (\Delta \rho g))^{1/2}$ between 
membrane and wall, where $\dot{\gamma}$ is
an effective shear rate, $R_0$ a mean vesicle radius, $\eta$ the 
fluid viscosity, and $g$ the gravitational acceleration. 
For $R_0=10\mu$m, $\dot{\gamma}=1$s$^{-1}$ (corresponding to a mean flow 
velocity of about $20 \mu$m/s) and the viscosity of water, we obtain the 
order-of-magnitude estimate of $\ell=10\mu$m, comparable to the 
channel size $L_z$. 
The displacement from the center implies an asymmetric deformation of 
the vesicle shape in glucose solution under flow.

\section{Theory for quasi-spherical vesicles in modulated channels}

We consider vesicles which have the same viscosity $\eta$ 
of the outer and inner fluids. Vesicles have constant volume $V$ and 
constant surface area $S$, so that the reduced volume $V^*$ and the 
excess area $\Delta_{\rm S}$ are defined by
$V^*= (R_{\rm V}/R_{\rm S})^3= (1+\Delta_{\rm S}/4\pi)^{-3/2}$ and
$\Delta_{\rm S}=S/R_{\rm V}^2 -4\pi$,
where $R_{\rm V}= (3V/4\pi)^{1/3}$ and $R_{\rm S}=(S/4\pi)^{1/2}$.

First, we derive an analytical description of the flow behavior 
of small quasi-spherical vesicles with $\Delta_{\rm S} \ll 1$,
based on the theory for linear flows~\cite{seif99,misb06,nogu07b,lebe08,turi08}.
The microchannel has a constant height $L_z$ and a periodically modulated  
width with walls at $\pm h_y(x)$, where
\begin{equation}
h_y(x)= \frac{L_y}{2}\Big\{1 +a_y \cos\Big(\frac{2\pi x}{L_x}\Big)\Big\}.
\end{equation}
Here, $L_x$ is the periodicity length along the channel (see Fig.~\ref{fig:setup}).
The flow field ${\bf v}({\bf r})$ for weakly modulated channels
($a_y \ll L_x/L_y$) is obtained from lubrication theory,
\begin{eqnarray}
v_x &=& (9v_{\rm m}L_y/8h_y)\{1-(y/h_y)^2\}\{1-(2z/L_z)^2\}, \nonumber \\
v_y &=& -\pi v_x  y L_y a_y\sin(2\pi x/L_x)/{h_y}L_x, \\
v_z &=& 0, \nonumber
\end{eqnarray}
where $v_{\rm m}$ is the mean flow velocity.
Furthermore, we consider small vesicles with $R_{\rm V} \ll L_y, L_z$,
and neglect backflow effects due to the channel wall.

The vesicle shape is expanded in spherical coordinates as 
$r= R_{\rm V}(1+\sum_{l,m}u_{l,m}Y_{l,m})$, with 
the spherical harmonics $Y_{l,m}(\theta,\varphi)$ and the 
polar axis in the $z$ direction.
We consider a vesicle shape which is mirror symmetric in the $z$ direction,
which implies, {\it e.g.}, $u_{2,\pm 1}=u_{3,0}=u_{3,\pm 2}=0$.
The curvature energy of the membrane with bending rigidity $\kappa$ and 
reduced surface tension $\sigma$ is given by
\begin{equation}
F = (\kappa/2)\sum_{l,m} E_l |u_{l,m}|^2 + \kappa(12+2\sigma/3)u_2^{(3)},
\end{equation}
where $E_l=(l+2)(l-1)\{l(l+1)+ \sigma\}$. Here, the third order term 
$u_2^{(3)}=(\sqrt{5/\pi}/7)({u_{2,0}}^3- 6u_{2,0}|u_{2,2}|^2)$ 
for $l=2$ is necessary to obtain a prolate shape as thermal-equilibrium 
state \cite{miln87,lebe08}.
With the Stokes approximation and the Lamb solution for the flow field, 
the dynamics of the amplitudes $u_{l,m}$ of a vesicle moving 
along the center line ($y=z=0$) of the channel with velocity 
${\bf v}(x_{\rm G},0,0)=(9v_{\rm m}L_y/8h_y,0,0)$ is governed by
\cite{seif99,lebe08,turi08}
\begin{equation}
\label{eq:sphdy}
\frac{\partial u_{l,m}}{\partial t} = s_{l,m}(x_{\rm G}) - 
                                      \kappa^* \Gamma_l f_{l,m}
\end{equation}
with reduced bending rigidity 
$\kappa^*=\kappa L_y/(\eta {R_{\rm V}}^3v_{\rm m}) = L_y/(v_{\rm m}\tau)$
and $\Gamma_l = l(l+1)/(2l+1)(2l^2+2l-1)$. Here, $\tau=\eta R_{\rm V}^3/\kappa$ is
the characteristic relaxation time of a vesicle. Two kinds of forces determine
the membrane deformation in Eq.~(\ref{eq:sphdy}), the curvature force 
$f_{l,m}= E_l u_{l,m} - (12+2\sigma/3)\partial u_2^{(3)}/\partial u_{l,m}$, 
and the force $s_{l,m}(x_{\rm G})$
due to the elongational flow.
The force $s_{l,m}(x_{\rm G})$ is given by 
$s_{l,m}(x_{\rm G})=( a_{l,m}+B_l b_{l,m})L_y/R_{\rm V}v_{\rm m}$ 
\cite{seif99,lebe08,turi08}.
Here, $B_l=1/(2l^2+2l-1)$, and $a_{l,m}$ and $b_{l,m}$ are the spherical
harmonics expansion of 
$v_r({\bf r})$ and $R_{\rm V}\partial v_r({\bf r})/\partial r$ 
at $|{\bf r}-{\bf r}_{\rm G}| = R_{\rm V}$, respectively,
where $v_r({\bf r})$ is the radial components of the flow field ${\bf v}({\bf r})$ 
in the absence of the vesicle.
The tension $\sigma$ is determined by the area constraint 
$\partial \Delta_{\rm S}/\partial t=0$ with
$\Delta_{\rm S}=\sum_{l,m} (l+2)(l-1)|u_{l,m}|^2/2 - 2u_2^{(3)}/3$.
We consider a vesicle with a small excess area $\Delta_{\rm S}=0.1$ 
(corresponding to $V^*=0.988$), 
where it is sufficient to take into account $l=2,3$ modes, 
and a channel with $L_y=L_z=20R_{\rm V}$ and various $L_x$ and $a_y$.

\begin{figure}
\includegraphics{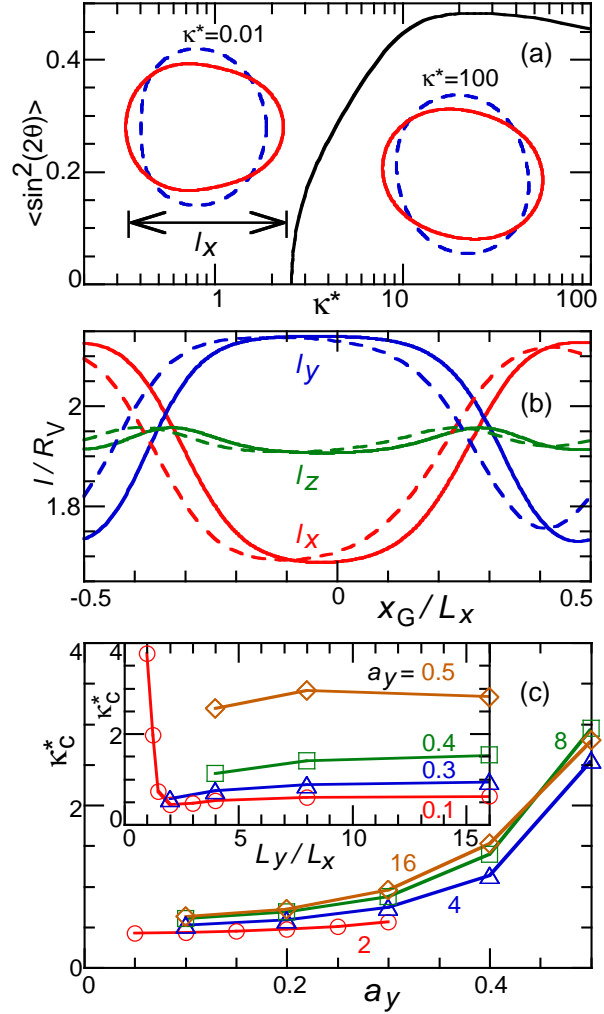}
\caption{  \label{fig:sph}
(Color online)
Dynamics of a quasi-spherical vesicle at $V^*=0.988$.
(a) Dependence of the tilt angle $\theta$ of a vesicle
on $\kappa^*$, for $L_x/L_y=4$ and $a_y=0.5$.
Here, $\langle \sin^2(2\theta) \rangle = 
\langle ({\rm Im}(u_{22})/|u_{22}|)^2 \rangle$
describes the deviation from symmetric shape (with 
respect to the $xz$ plane).
The insets show sliced snapshots of vesicles in the $xy$ plane 
 for $\kappa^*=0.01$ and $\kappa^*=50$.
Solid and dashed lines indicate shapes of extremal elongation or tilt.
(b) Maximum vesicle lengths in the $x,y,z$ directions as a function of the 
center-of-mass position $x_{\rm G}$, for $\kappa^*=0.01$ and $a_y=0.5$.
Solid and dashed lines represent results for $L_x/L_y=4$ and $L_x/L_y=16$, 
respectively.
(c) Critical reduced bending rigidity  $\kappa^*_{\rm c}$ as a function 
of the corrugation amplitudes $a_y$ for
channel geometry $L_x/L_y=2$, $4$, $8$, and $16$.
The inset shows  $\kappa^*_{\rm c}$ as a function 
of $L_x/L_y$ for $a_y=0.1$, $0.3$, $0.4$, and $0.5$.
}
\end{figure}

In fast flow, where flow forces dominate and the reduced bending rigidity is
small, $\kappa^* \ll 1$, the analysis of Eq.~(\ref{eq:sphdy}) shows that the shape is 
mirror symmetric with respect to 
both $xz$ and $xy$ planes, and that the vesicle lengths $l_x$ and $l_y$ oscillate
(see Fig.~\ref{fig:sph}).
The flow elongates the vesicle in the $y$ and $x$ direction
for $-L_x/2<x< 0$ and $0<x< L_x/2$, respectively.
In the regime $\kappa^* \ll 1$, the extremal elongations
are determined by the balance of the flow forces, $s_{l,m}(x_{\rm G})$,
and the area constraint.  The forces $|s_{2,2}(x_{\rm G})|$ and 
$|s_{3,3}(x_{\rm G})|$ have maxima at $x_{\rm G}=\pm 0.38L_x$ and 
$x_{\rm G}=\pm L_x/2$ for $a_y=0.5$, respectively.
The resulting positions of extremal elongations are found to be
close to $x_{\rm G}=0$ and $x_{\rm G}=L_x/2$ (see Fig.~\ref{fig:sph}(b)).

In contrast, in slow flow with $\kappa^* \gg 1$, the vesicle is predicted not 
to change its shape, but to display a periodic oscillation of its orientation.
The tilt angle $\theta$ oscillates around $\pi/4$ 
(or $-\pi/4$ depending on initial positions), see snapshot in Fig.~\ref{fig:sph}(a).
Thus, a symmetry breaking occurs with decreasing flow velocity.

Both types of vesicle motions are limit cycles.
A sharp but continuous transition between these two motions occurs 
at a critical reduced bending rigidity $\kappa^*_{\rm c}$, where the 
tilt angle $\theta$ vanishes (see Fig.~\ref{fig:sph}(a)).
The critical value $\kappa^*_{\rm c}$ increases with increasing corrugation
$a_y$ of the channel, but is almost independent of $L_x$ for $L_x/L_y > 2$
(see Fig.~\ref{fig:sph}(c)).
The symmetric vesicle deformation reduces the disturbance of the original 
flow but increases the free energy of a vesicle.
At small or large $\kappa^*$, the former or latter contribution dominates, 
respectively. 

\begin{figure}
\includegraphics[width=0.9\linewidth]{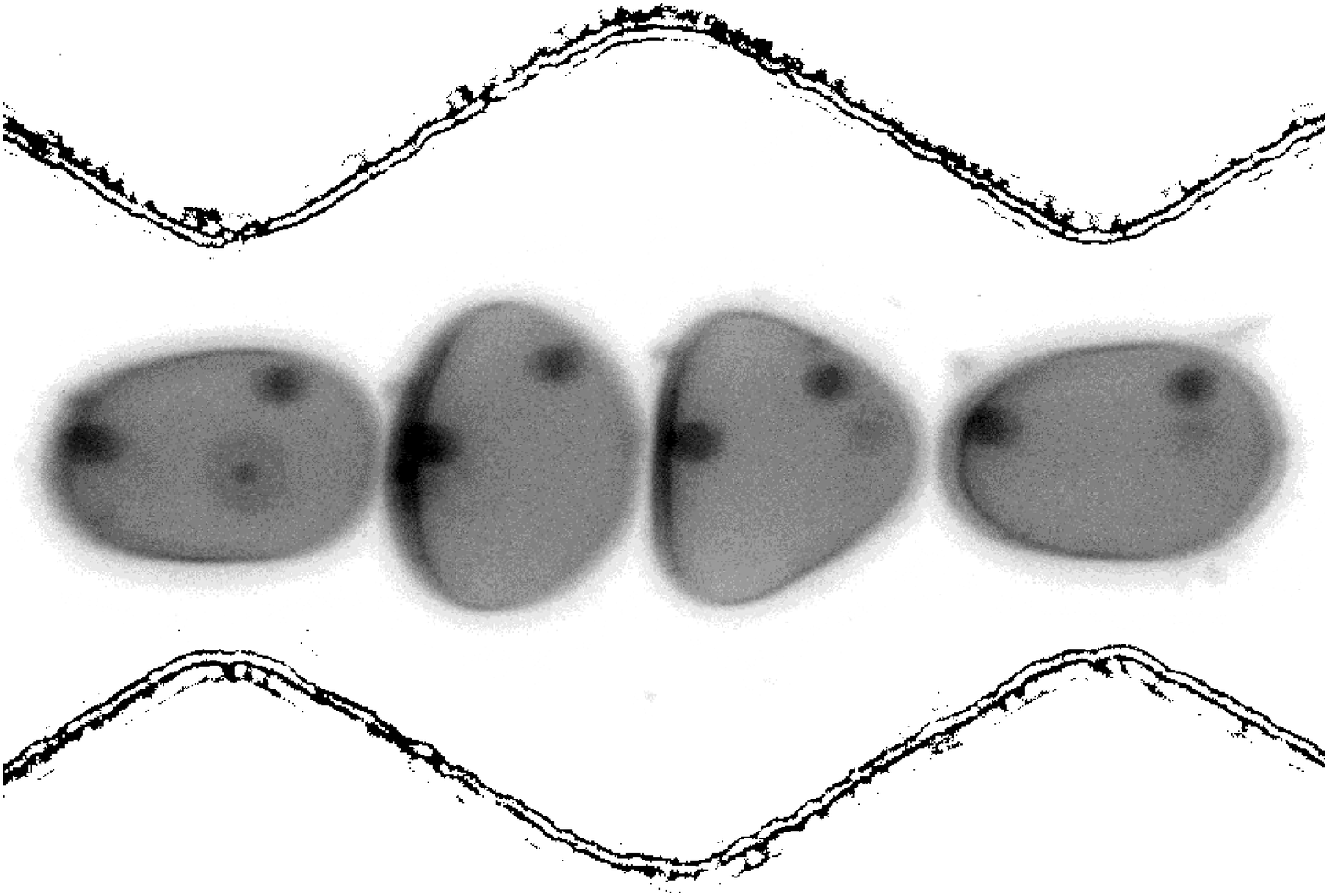}
\includegraphics[width=0.9\linewidth]{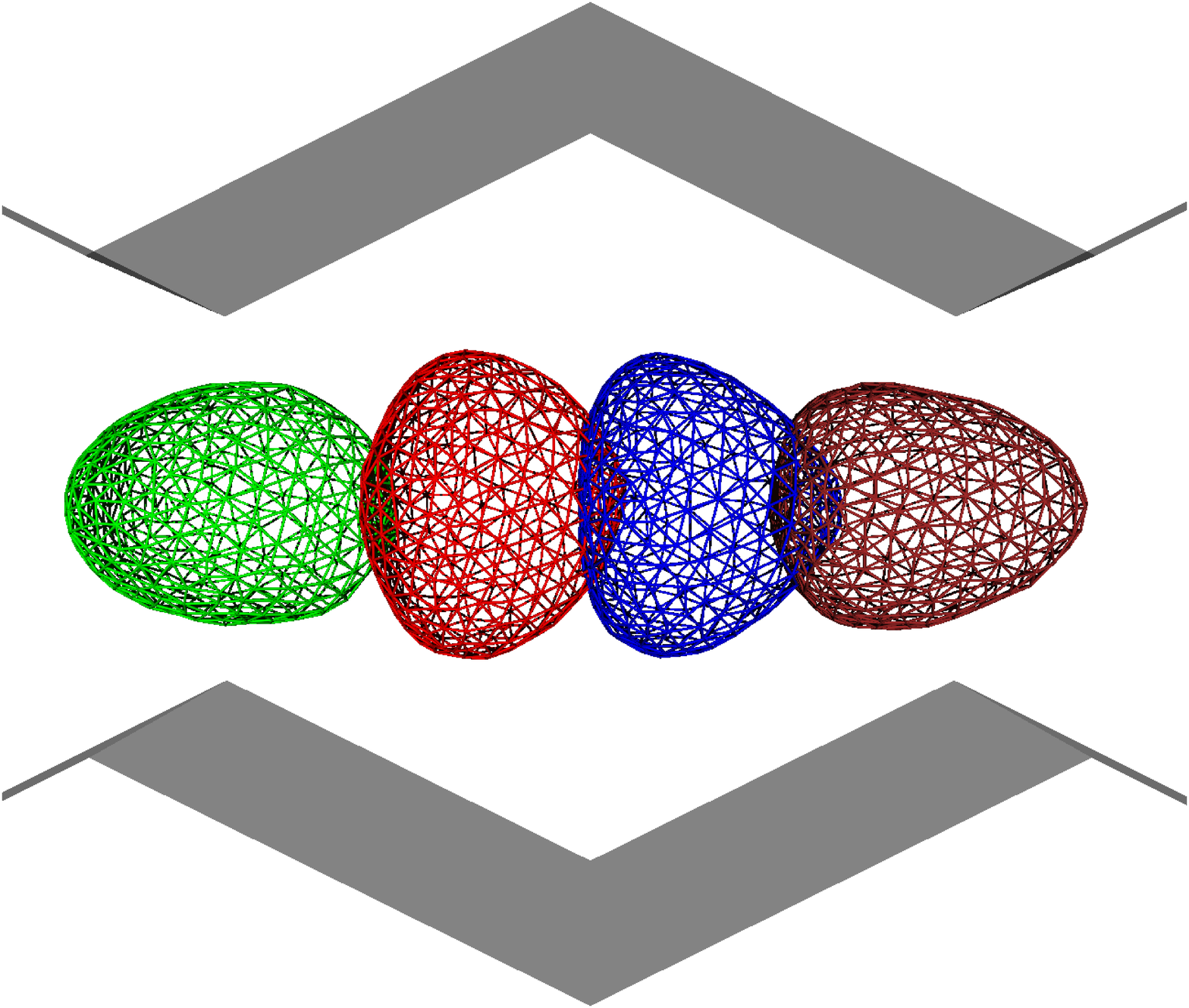}
\caption{  \label{fig:overlay}
(Color online)
Sequence of snapshots (at equal time intervals) of vesicles 
with $V*=0.96$ and $R_{\rm S}/L_y=0.21$
moving through a microchannel, experimentally observed by optical 
microscopy (in sucrose solution) for $\kappa^*=0.04$ 
($v_{\rm m}=38 \mu\rm{m}/\rm{sec}$, $L_y=75\mu$m) (top), 
and in simulation for $\kappa^*=0.08$ (bottom).  
For the experimental data, the time interval between images is 0.97~sec.}
\end{figure}

\section{Mesoscale Hydrodynamics Simulations of Vesicles under Flow}

We employ mesoscale hydrodynamics simulations to study the flow behavior
of vesicles in narrow channels.
A dynamically-triangulated surface model for the membrane \cite{gg:gomp97f} 
is combined with a particle-based mesoscale simulation technique ---  
multi-particle collision dynamics (MPC)  
\cite{male99,kapr08,gomp09} --- for the embedding fluid.
A detailed description of this approach to model vesicles dynamics under 
flow can be found in Refs.~\cite{nogu05,nogu05b}.
In MPC, the fluid is described by $N_{\rm s}$ point-like
particles of mass $m_{\rm s}$. After free streaming of every particle 
for a time step $\Delta t$, the fluid particles collide in cubic boxes 
of lattice constant $a=L_x/16$, which is
also the mean length between membrane vertices. We use the fluid 
viscosity $\eta = 550 \sqrt{m_{\rm s}k_{\rm B}T}/a^2$ (with number density
$n=100a^{-3}$ and the time step $\Delta t = 0.01a\sqrt{m_{\rm s}/k_{\rm B}T}$),
corresponding to low Reynolds numbers, to simulate experimental conditions.
We study vesicles with bending rigidity
$\kappa = 20k_{\rm B}T$ where $k_{\rm B}T$ is the thermal energy.
The flow velocity is chosen to be $v_{\rm m}\tau/L_y = 170 (R_{\rm V}/L_y)^3$ 
for a channel with $L_y=L_z=L_x/2$,
corresponding to a slow vesicle relaxation compared to the passage time 
through a channel segment. For vesicles with $R_{\rm S}/L_y = 0.36$ 
and $R_{\rm S}/L_y = 0.46$ at $V^*=0.9$,
the reduced bending rigidity is $\kappa^* = 0.14$ and $\kappa^* = 0.07$, 
respectively.
In this regime, the vesicle dynamics is not very sensitive to $\kappa^*$.

\begin{figure}
\includegraphics{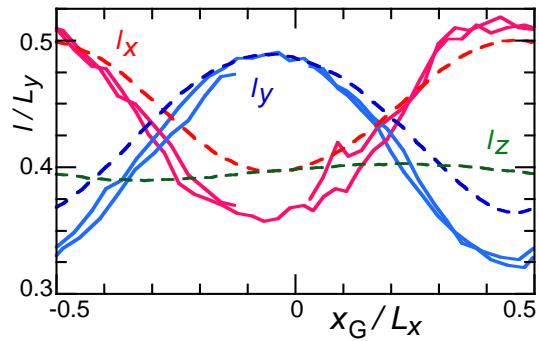}
\caption{  \label{fig:lmax}
(Color online)
Maximum lengths $l_x$, $l_y$, and $l_z$ of vesicles in experiments 
(solid lines) and simulations (dashed lines) from the same data as 
shown in Fig.~\ref{fig:overlay}.
}
\end{figure}

\section{Vesicles in sawtooth-shaped channels in fast flows}

We investigate the dynamics of large prolate vesicles in fast 
flows, corresponding to small $\kappa^*$, both by experiments and simulations. 
In this case, hydrodynamic interactions between the vesicle and wall are not 
negligible. We consider now a periodically-patterned microchannel with 
a saw-tooth shape (see Figs.~\ref{fig:setup} and \ref{fig:overlay}), for which
the walls are located at  
\begin{equation}
h_y(x)= \frac{L_y}{2}\Big\{ 1 +  a_y\Big(1 \pm \frac{4x}{L_x}\Big)\Big\}
\end{equation}
for $-L_x/2<x\leq 0$ and $0<x\leq L_x/2$, respectively. 
Two types of channels are used, which are characterized by
$L_y=50\mu$m, $a_y=0.2$ and $L_y=75\mu$m, $a_y=0.33$
(as well as $L_z=50\mu$m and $L_x=100\mu$m). 
The glucose and sucrose solutions are used for the experiments with 
the narrower and wider channels, respectively. 
Typical experimental flow velocities are in the 
range $v_{\rm {m}}=10$ -- $100\mu$m/s.

First, we describe the dynamics in the wider channel ($L_y=75\mu$m).
Experimental and simulation results for the shape deformation are shown in 
Figs.~\ref{fig:overlay} and \ref{fig:lmax}.
As expected, the vesicles change their shapes periodically. 
For $V^* \gtrsim 0.9$, vesicle shapes vary {\it continuously} between an 
almost ellipsoidal  (in the narrow part of the channel) 
and a cone-like (in the wide part of the channel) bullet shape, 
see Fig.~\ref{fig:overlay}.  This is in contrast to the behavior 
for {\it unbounded} steady Poiseuille flow, where perturbation theory predicts 
a coexistence between bullet- and parachute-like shapes at the center line 
\cite{misbah2009}.
The amplitudes and phases of the deformations in the $x$- and 
$y$-directions (see Fig.~\ref{fig:lmax}) agree well
between simulations and experiments, while very little 
deformation is seen in the $z$-direction.
The small difference of amplitudes between the experiment 
and simulation is probably due to the somewhat different values of the 
reduced volume and radius of the vesicle, since an accurate experimental 
estimation is difficult 
(because the length in the $z$ direction could not be measured).
The dynamics also agrees qualitatively  with the theoretical predictions  
for quasi-spherical vesicles, 
compare Fig.~\ref{fig:sph}, although the parameters are outside the range
of validity of the theoretical approximations.

\begin{figure}
\includegraphics{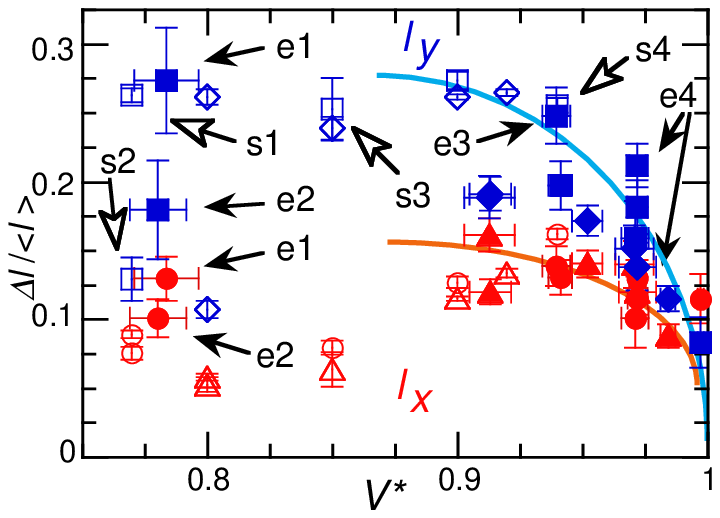}
\includegraphics{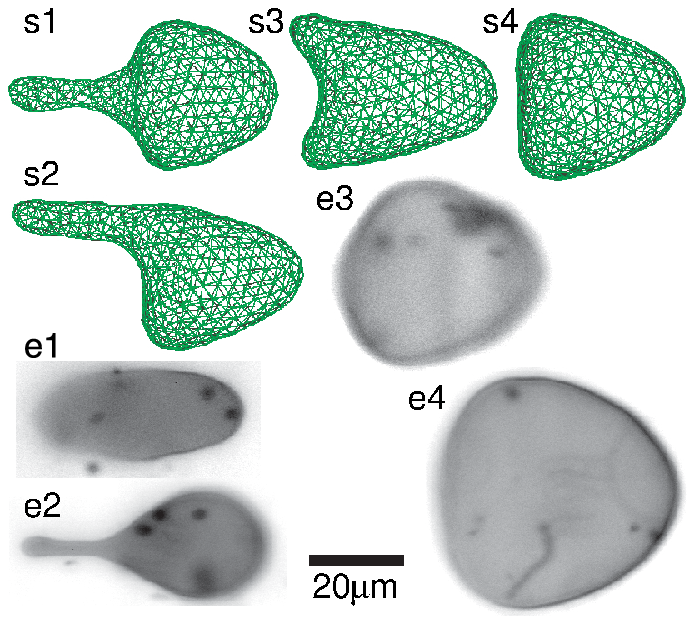}
\caption{  \label{fig:mean}
(Color online)
Normalized oscillation amplitudes $\Delta l/\langle l \rangle$ in the
$x$ (red) and $y$ (blue) directions. 
Open symbols represent simulation data for $R_{\rm S}/L_y=0.36$ 
($\circ$, $\Box$) and $R_{\rm S}/L_y=0.46$ 
($\triangle$, $\diamond$) at $a_y=0.2$, respectively.
Closed symbols represent experimental data for $L_y=50\mu$m with 
$a_y=0.2$ in glucose solution ($\bullet$, $\blacksquare$) and $L_y=75\mu$m with $a_y=0.33$ 
 in sucrose solution ($\blacktriangle$, $\blacklozenge$), respectively.
The solid lines are guide to the eye.
The lower part shows snapshots of simulations (s1, s2, s3, s4)
and experiments (e1, e2, e3, e4) for $L_y=50\mu$m.
All snapshots are displayed with the same length scale. 
The pictures e1 and e2 show the same vesicle before and after the
transformation from ellipsoid to tailed shape, respectively. 
All images (e1-e4) are vesicles in glucose solution. 
}
\end{figure}

At reduced volumes $V^* \lesssim 0.9$, we find that the vesicles 
deform into novel shapes, while they are still ellipsoidal in the 
absence of flow.
At $V^* \sim 0.85$, the simulated vesicles develop two tails in the $xy$ plane 
(see Fig.~\ref{fig:mean}(s3)).
This tail position is very stable; the tails quickly return to the $xy$ plane,
even from an initial state with the tails in $xz$ plane (obtained by $\pi/2$ rotation).
At $V^* \sim 0.80$, the symmetric tails become unstable; they are replaced by 
a stable shape with a single asymmetric tail (see Fig.~\ref{fig:mean}(s2)).
In this case, the configuration with a tail in the $xz$ plane  
is also stable within the accessible simulation time (see 
Fig.~\ref{fig:mean}(s1)).
A single-tail shape is also observed in our experiments 
with vesicles in glucose solution (see Fig.~\ref{fig:mean}(e2));
here, the off-center motion induced by the buoyancy force implies 
that the tail is found in the $z$ direction.
Recently, Abkarian {\it et al.} \cite{abka08} found a 
similar asymmetric tail for RBCs in very fast flows in 
homogeneous capillaries
(with flow velocity $v_m = 10-35$~cm/s and 
capillary diameter $10 \mu$m). 
Such a long tail is much more difficult to form in RBCs due to the 
shear resistance of the spectrin network, and 
may even require a local 
separation of the lipid bilayer from the spectrin network.
Thus, the formation of an asymmetric long tail is a generic 
feature in capillary flows. Structured channels promote
tail formation already at smaller flow velocities. 
The low contrast and the blurry contour in the rear part of the vesicles 
in the experimental micrographs of Fig.~\ref{fig:mean} (e3, e4) indicates
that vesicles exhibit enhanced shape fluctuations at the rear part,
due to a locally reduced membrane tension; 
the low contrast regions in the shapes of Fig.~\ref{fig:mean} (e1, e2) 
is probably due to a tilt of the vesicle axis and a 
vesicle asymmetry induced by off-center motion with a buoyancy force.

\begin{figure}
\includegraphics[width=0.95\columnwidth]{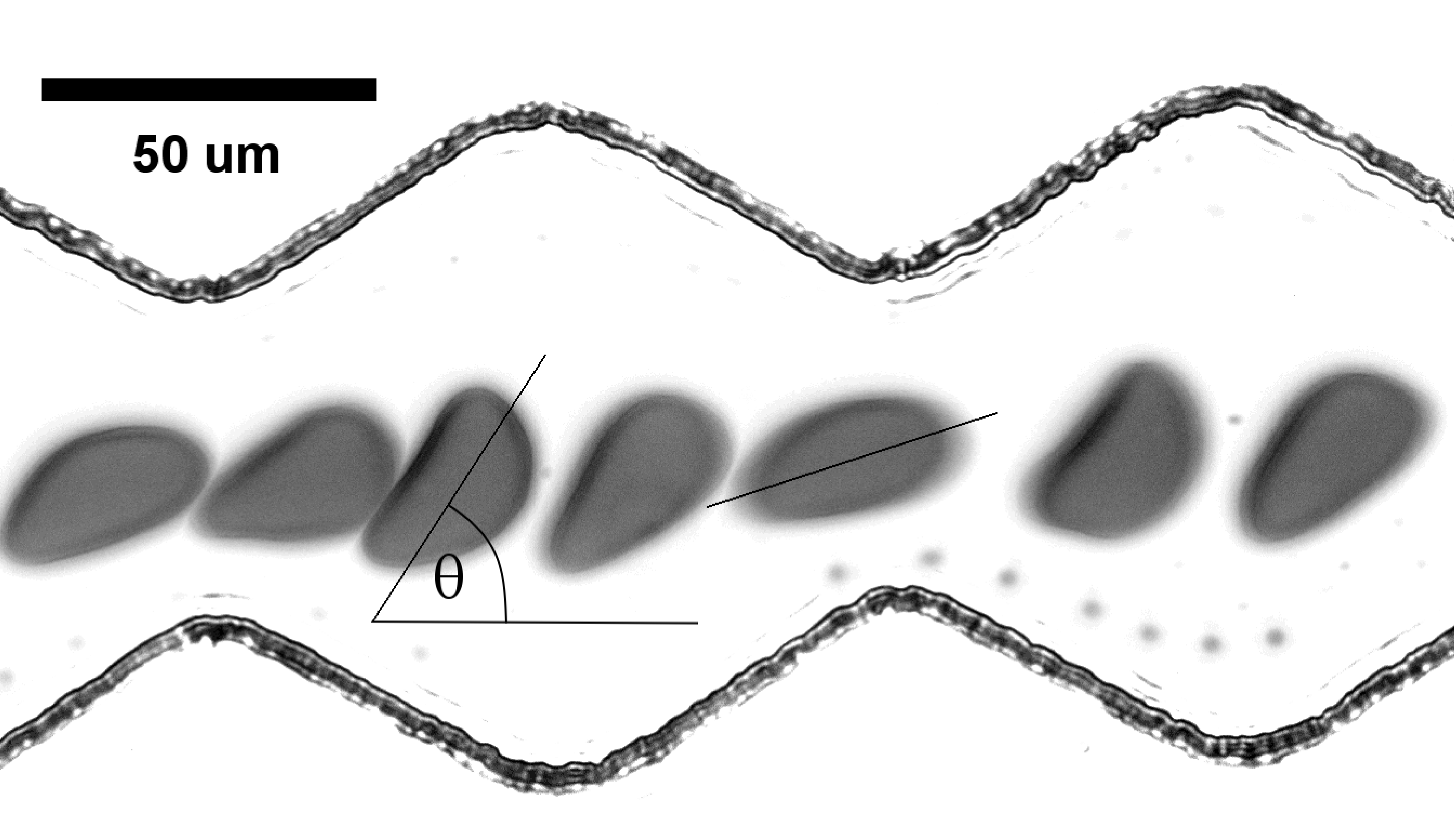}
\includegraphics{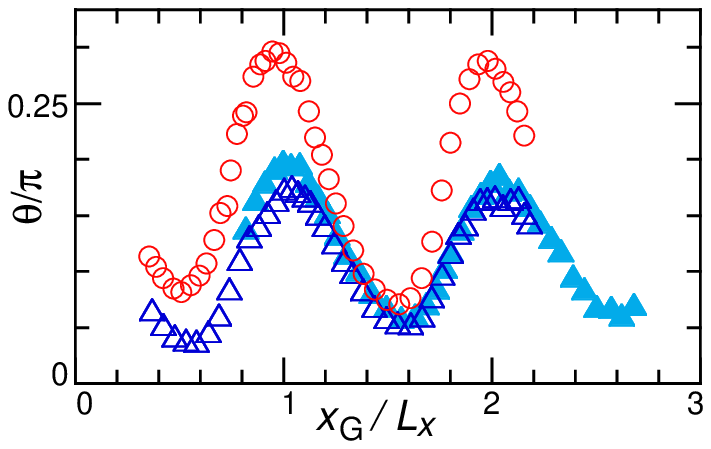}
\caption{  \label{fig:sph2}
(Color online)
Oriental oscillation of a vesicle (in sucrose solution) in a wide 
structured channel with $L_y=75\mu$m. 
(a) Micrograph of an oscillating vesicle with $R_{\rm {V}}=24 \mu$m, 
$v=34$ $\mu$m/s, corresponding to $\kappa^*=0.0132$.
The image is a superposition of video frames taken at time intervals of $1$s. 
(b) Orientational angle as a function of the center-of-mass position 
$x_{\rm G}$.
The (red) circles ($\circ$) represent the same data as in (a).
The vesicle oscillates between the extrema of $\theta_{\rm {min}}=0.07\pi$ 
and $\theta_{\rm {max}}=0.3\pi$.
The (blue) triangles represent another vesicle 
(with $R_{\rm {V}}=18\mu$m, $v=\pm 55\mu$m/s).
Here, we have inverted the direction of the fluid flow; symbols
indicate angles before ($\triangle$) and after ($\blacktriangle$) 
flow inversion. 
}
\end{figure}

Our experimental and simulation results for the oscillation 
amplitude $\Delta l$, displayed in top part of Fig.~\ref{fig:mean}, show 
that the 
difference between maximum and minimum of $l_x$ or $l_y$ is quite 
insensitive to the reduced volume $V^*$ and the vesicle size $R_{\rm S}$. 
In the experiments, $V^*$ is estimated
from the prolate ellipsoid shape with the lengths $\langle l_x \rangle$ 
and $\langle l_y \rangle$ of the long and the two short axes, respectively.
The finite amplitudes for $V^*\simeq 1$ are caused by errors in the 
estimation of $V^*$.
The velocity $v_{\rm {ves}}$ and average size 
$l_{\rm m}= (\langle l_x \rangle + \langle l_x \rangle)/2 $ of liposomes 
are varied in experiments in the ranges
$v_{\rm {ves}}=14$ -- $45\mu$m/s  and $l_{\rm m}= 20$ -- $40\mu$m,
respectively.
The amplitude $\Delta l_y$ of the vesicle with asymmetric tail in the 
$xy$ plane (s2) is smaller than in $xz$ plane (s1), 
since its tail is still at the wide part of the channel and elongated 
when the main body is at its narrowest part.

For vesicles, which flow in the center of the channel,
similar symmetric shape oscillations are observed in the
narrow and wide channels (in agreement with our predictions 
for fast flows from perturbation theory).  However, when the vesicle is 
slightly displaced from the center ($y=0$) in the wider channel
it displays orientational oscillations, as shown in Fig.~\ref{fig:sph2}.
Here, the asymmetry of the flow field acts to magnify
the orientational oscillations, which are accompanied by strong 
shape oscillations between slipper and ellipsoidal shapes.
Thus, a mixed state of orientational {\it and} shape 
oscillations appears for $\kappa^* \lesssim 1$; it  
appears to be long-lived due to the small lift force for wider channels.

\section{Vesicles in sawtooth-shaped channels in slow flows}

Finally, we want to briefly discuss the dynamics of large vesicles
at slow flows, with $\kappa^* \gg 1$.
If the vesicle size is larger than the smallest channel width 
$L_y^{\rm {min}}$, vesicles can  not flow through the channel without 
a shape deformation.  Thus, with decreasing flow velocity,
the dynamics changes from shape oscillations 
to a trapped state \cite{gomp95} (instead of the orientational 
oscillation). 
The transition velocity depends on $R_{\rm S}/L_y^{\rm {min}}$,
$V^*$, and $\kappa^*$.
A microchannel with a strong variation of channel width 
(for instance the channel in Fig.~6(a) of Ref.~\cite{abka08}), 
is suitable for a study of this trapped-escape transition.

\section{Summary}

We have studied the dynamics of vesicles in structured microchannels,
both theoretically and experimentally. 
For large reduced volumes, the vesicles periodically 
change their shape (large flow rates) or their 
orientation (small flow rates),
depending also on the channel geometry.  
For smaller reduced volumes, we find a novel shape with 
a long asymmetric tail.
The good agreement of theoretical and experimental results  
shows that flow of vesicle suspensions in complex 
flow geometries can now be understood quantitatively.

\acknowledgments
We acknowledge support of this work through the DFG priority program 
``Nano- and Microfluidics'' (SPP 1164) and the 
``Nanosystems Initiative Munich'' (NIM).

\end{document}